\documentclass[onecolumn]{aa}
\usepackage{natbib}
\usepackage{graphicx}
\bibpunct{(}{)}{;}{a}{}{,} % to follow the A&A style

\begin{document}

\title{A method to simulate inhomogeneously irradiated objects with a superposition of 1D models}

\author{H.~M. G\"unther\inst{1,2} \and A. C. Wawrzyn\inst{1}}

%\offprints{H.~M. G\"unther,\\ \email{moritz.guenther@hs.uni-hamburg.de}}
\institute{Hamburger Sternwarte, Universit\"at Hamburg, Gojenbergsweg 112, 21029 Hamburg, Germany\and 
Harvard-Smithsonian Center for Astrophysics, 60 Garden Street, Cambridge, MA 02138, USA\\ \email{moritz.guenther@hs.uni-hamburg.de}}
\date{Received TBD / Accepted TBD}
\abstract{In close binary systems the atmosphere of one or both components can be significantly influenced by irradiation from the companion. Often the irradiated atmosphere is simulated with a single-temperature approximation for the entire half-sphere.}{We present a scheme to take the varying irradiation angle into account by combining several separate 1D models. This is independent of the actual code which provides the separate stellar spectra.}{We calculate the projected area of zones with given irradiation angle and use this geometrical factor to scale separate 1D models. As an example we calculate two different irradiation scenarios with the PHOENIX code.}{The scheme to calculate the projected area is applicable independent of the physical mechanism that forms these zones %\footnote{We provide an IDL code for these computations in electronic form only at the CDS via anonymous ftp to cdsarc.u-strasbg.fr (130.79.128.5) or via http://cdsweb.u-strasbg.fr/cgi-bin/qcat?J/A+A/2010/14824}
. In the case of irradiation by a primary with T=$125\;000$~K, the secondary forms ions at different ionisation states for different irradiation angles. No single irradiation angle exists which provides an accurate description of the spectrum. We show a similar simulation for weaker irradiation, where the profile of the H$\alpha$ line depends on the irradiation angle.}{}
\titlerunning{1.5D model atmospheres}
\maketitle

\section{Introduction}
\label{introduction}
The physical conditions in a range of objects e.g. hot exoplanets and close binaries are fundamentally influenced by external irradiation. Often the central source is much hotter than its companion, so that the primary can be modelled as stand-alone, but the irradiated secondary has an atmospheric structure which differs markedly from the unirradiated case. At the substellar point, where the primary is closest and the irradiation is perpendicular to the secondary's surface a hot spot develops. For different latitudes the intensities and irradiation angles vary, leading to zones of different temperatures on the secondary; in an ideal case the hot spot coincides with the sub-stellar point and rings of constant temperatures form around it. Observations of cataclysmic variables (CVs) \citep{1992MNRAS.257..476D} and exoplanets \citep{2007Natur.447..183K} indicate, that the hot spot may be displaced from the substellar point in some cases.

An approximation for the visibility of the irradiating star (as a point source) on the surface of the irradiated one was already presented by \citet{1926MNRAS..86..320E}. This is described in more detail by \citet[][see his Sect. IV.6]{1959cbs..book.....K}. \citet{1968Ap&SS...2...61N}, however, has shown that these approximations are in error not only in the 'penumbral' (a partial shadow, as in an eclipse, between regions of complete shadow and complete illumination) regions, but also at full phase, which makes spherical treatment of the source necessary and he provides a scheme for it. Later \citet{1973MNRAS.164...53W} presented an approximation for the integrated bolometric flux, which is specifically designed for computer calculations, though his results are similar to the ones of \citet{1959cbs..book.....K}. Simulations of irradiated stellar atmospheres started more than two decades ago, using opacities for a grey atmosphere, and have improved since \citep{1985A&A...147..281V,1987PASJ...39..163N,1990A&A...228..231N,1993MNRAS.264..641B}. These improvements include metalicity effects \citep{1992A&A...256..572C}, a treatment of polarised light by \citet{1993MNRAS.263..989C}, non-LTE effects \citep{1996ApJ...471..930P,2000ASPC..214..705K} and line profiles for non-static atmospheres \citep{1998A&AS..132...45P} as well as limb-darkening and gravity-brightening \citep{1997A&A...326..257A,1999A&AS..135..555A,1999A&A...346..556A,2007A&A...470.1099C}. Nowadays the existing models can be used to fit parameters of individual systems \citep{2009A&A...505..227W}. \citet{1993MNRAS.264..641B} introduced the idea of entropy-matching in the convections zone, which presumably is well-mixed and thus can transport heat horizontally.

A treatment of irradiation is implemented in the radiation transfer code PHOENIX \citep{1999JCoAM.109...41H} with the proper geometric factors, which are needed for extended primaries leading to different visibilities and irradiation angles on any given point on the secondary \citep{2004ApJ...614..338B,2005ApJ...632.1132B}. As long as full 3D calculations are out of reach of the current computer capabilities for state of the art atmosphere modelling codes, we need a method to combine multiple 1D simulations each calculated for an atmospheric structure appropriate for a given temperature and irradiation angle. This requires an integration over the inhomogeneous stellar surface of the secondary, were regions of different temperature are seen under a different angle. We were surprised not to find an extensive description of the geometry for close binaries in the literature; only \citet{2003ARep...47..540L} treats a geometrically similar problem in the context of accreting young stars. The calculations are conceptionally simple, although somewhat tedious. Often this task is not explicitly described \citep{2008AstL...34..423S} or referred to numeric methods. \citet{2005ApJ...632.1132B} present spectra for irradiated planets, where the dayside is modelled in ten rings. To obtain the weighting factor, they use a statistical method and distribute about 2000 points on the planetary surface. A surface area is than assigned to each point and the integral is performed as the sum over all visible points \citep[see also][]{1996JQSRT..56...97S}. In contrast, we want to proceed analytically as far a possible. In Sect.~\ref{geometry} we provide ready-to-use formulae and associated IDL code to calculate the proper weighting of spectra calculated for different positions on the star. Using this method we simulate two exemplary spectra of irradiated close binaries with stellar parameters appropriate for pre-CVs with a hot sdO and analyse which new spectral features are revealed when the heated side is represented by a superposition of different spectra (Sect.~\ref{star15d}). We end with a short conclusion in Sect.~\ref{conclusion}.

\section{The geometry}
\label{geometry}
\subsection{Model assumptions}
The calculations we present in this section are purely geometrical. We calculate the projected surface area on the plane of the sky of a patch on a sphere, given the limiting longitude and latitude. The problem is complicated by the fact that we may view the sphere under an arbitrary angle and that parts may be eclipsed by the primary star, which is also assumed to be spherical.

The physical situation we have in mind is a close binary system, e.g. a pre-CV or a hot Jupiter. The orbits may be eccentric, as long as the ephemerides or at least the separation at the time in question is known. 

The formulae in this section are independent of the physical mechanism that produces the spectral differences on the observed body, they have to be taken into account by the spectral synthesis code. We think of rings of constant temperature, which develop mainly due to the irradiation angle. In principle winds are expected to smooth the temperature gradients, but they are difficult to model in 1D. In any case, the equations in this section can be used to add up different spectra from different patches on the star. Also, in principle, the emission can be time-depended, but in many systems this is not an issue. In circularised orbits with primaries of reasonably constant luminosities a stationary temperature distribution on the irradiated secondary will develop. If its rotation matches the orbital period, the hot spot will coincide with the sub-stellar point.

\subsection{The general problem}
\begin{figure}
\centering
\includegraphics[angle=90, width=5cm]{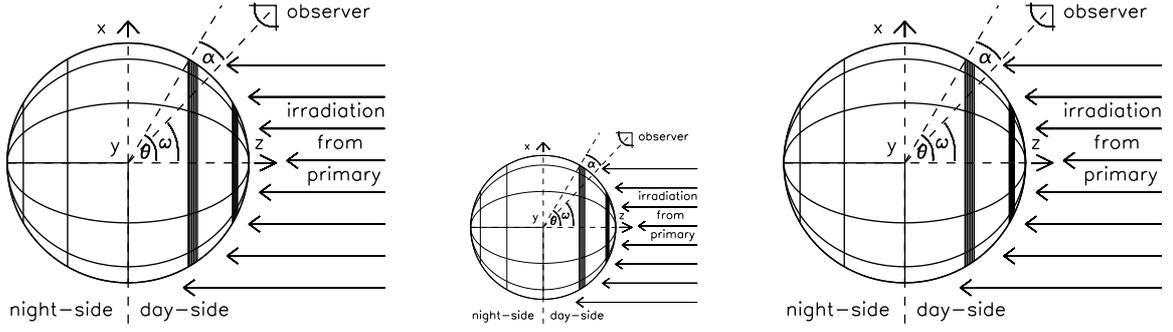}\hspace{1cm}
\includegraphics[angle=90, width=3cm]{14824fg1a.eps}\hspace{1cm}
\includegraphics[angle=90, width=5cm]{14824fg1a.eps}
\caption{\emph{left:} Side-view on the secondary; \emph{middle:} Some latitudes are fully visible, some partially visible and some invisible. \emph{right:} Latitudes close to grazing incidence are covered first. \label{fig:geom} }
\end{figure}

We adopt a standard spherical coordinate system with the origin at the centre of the secondary and the positive z-axis intersecting the centre of the primary: 
\begin{equation}
\label{r}
\vec{r}
=
\left(\begin{array}{c}
x \\
y \\
z
\end{array}\right)
= r
\left(\begin{array}{c}
\sin{\theta} \cos{\varphi} \\
\sin{\theta} \sin{\varphi} \\
\cos{\theta} \\
\end{array}\right)
\end{equation}
In this model lines of constant latitude $\theta$ (measured from the z-axis) receive the same amount of incident flux from the primary. Points near the terminator (which lies in the $z = 0$ plane) receive less incident flux than the substellar point because of shallower incident angles and larger distances from the primary. The problem is still rotationally symmetric around the $z$-axis, so we choose the observer to be in the $xz$-plane and call the angle between the $z$-axis and the line-of-sight (which is given by the unit vector $\vec{r}_{obs}$) $\omega$.
\begin{equation}
\vec{r}_{obs}
=
\left(\begin{array}{c}
x_{obs} \\
y_{obs} \\
z_{obs}
\end{array}\right)
=
\left(\begin{array}{c}
\sin{\omega} \\
0 \\
\cos{\omega} \\
\end{array}\right) \label{vectors}
\end{equation}
Figure~\ref{fig:geom} (left panel) shows a sketch of the setup.
% \begin{figure}
% \centering
% \includegraphics[angle=90, width=3cm]{z_sph1_coord_new.eps}
% \caption{Some latitudes are fully visible, some partially visible and some invisible. \label{fig:geomside} }
% \end{figure}
% 
% \begin{figure}
% \centering
% \includegraphics[angle=90, width=5cm]{z_sph1_coord_cover2.eps}
% \caption{Latitudes close to grazing incidence are covered first. \label{fig:cover} }
% \end{figure}

We normalise all distances to the radius of the irradiated secondary star ($r_{\mathrm{s}}=1$).
The angle $\alpha$ between the normal to the surface at any point of the surface (eqn.~\ref{r} with $r=1$) and the observer (eqn.~\ref{vectors}) is hence simply given by:

\begin{equation}
\cos{\alpha} = \sin{\theta}\cos{\varphi}\sin{\omega} + \cos{\theta}\cos{\omega} \label{alphaequation}
\end{equation}
If $\omega$ is known, e.g. for binaries with measured orbital parameters, this provides a relation $\varphi(\alpha)$ between the longitude $\varphi$ and the angle $\alpha$ for each latitude $\theta$:
%If the conditions are fulfilled we get our $\varphi(\alpha)$ simply from equation \ref{alphaequation}:
\begin{equation}
\varphi(\alpha)=\arccos\left(\frac{\cos\alpha-\cos\theta\cos\omega}{\sin\theta\sin\omega}\right)
\mathrm{ for } \;\theta \neq 0 \mathrm{ , } \;\omega \neq 0
\label{eqnvarphi}
\end{equation}
If the line-of-sight is parallel to the $z$-axis, i.e. $\omega=0$, then $\theta=\alpha$ for all $\varphi$.

We now proceed to calculate the surface area $A$ between the latitudes
$\theta_{1}$ and $\theta_{2}$, which is a ring on the surface of the star (Fig.~\ref{fig:geom}, left panel), projected on the observer's plane of the sky. We restrict the $A$ to those parts of the ring, that are seen under the angles between $\alpha_{1}$ and $\alpha_{2}$:
\begin{equation}
A(\theta_{1},\theta_{2},\alpha_{1},\alpha_{2})=\int^{\theta_{2}}_{\theta_{1}}
\int^{\varphi(\theta, \alpha_{2})}_{\varphi(\theta, \alpha_{1})} %r_{\mathrm{s}}^{2} ==1
 \sin{\theta} \cos\alpha(\varphi) \;
\mathrm{d}\varphi \; \mathrm{d}\theta \label{Asurface}
\end{equation}
The $\int \sin\theta \; \mathrm{d}\theta$ gives the surface area covered by the ring and the factor $\cos\alpha$ accounts for the projection on the sky.

The problem is assumed to be symmetric to the $xz$-plane so we
integrate only over one hemisphere ($\alpha \geq 0$) and multiply
the result by $2$.

If $\alpha$ boundaries are not explicitly given, we need to take into account that only parts of the ring are on the visible hemisphere (Fig.~\ref{fig:geom}, middle panel). Grazing views happen under $\alpha = -\frac{\pi}{2}$ or $+\frac{\pi}{2}$ and we need to distinguish between full, partial and not visible circles of latitude. Full circles only reach values of $\alpha$ in the range $|\omega-\theta|$ to $|\omega+\theta|$. Therefore, the integration boundaries for any `patch' to be observable are:
\begin{equation}
|\omega-\theta| \leq \alpha \leq \min(|\omega+\theta|,\pi/2)\;.
\end{equation}

In the general case the integral in eqn.~\ref{Asurface} cannot be solved analytically, but needs to be computed numerically.

We split the integral according to the visibility of the latitude in $A_{\mathrm{d}}$ containing fully visible circles on the day-side, $A_{\mathrm{p}}$ containing the partially visible circles and $A_{\mathrm{n}}$ for fully visible latitudes on the night-side:

\begin{equation} \label{Atotal}
A(\theta_1,\theta_2) = A_{\mathrm{d}}(\theta_1,\theta_2) + A_{\mathrm{p}}(\theta_1,\theta_2) + A_{\mathrm{n}}(\theta_1,\theta_2).
\end{equation}

%The boundary between visible and invisible parts is the angle
%$\alpha = -90$� and $+90$� respectively, where the horizon is
%touched only.

Angles $\theta$ belong to fully visible circles on the dayside for
\begin{equation}
0  \leq \theta \leq  \left|\frac{\pi}{2}-\omega\right| \textrm{ with } \omega<\frac{\pi}{2},  \label{intlimitsfull}
\end{equation}
on the nightside for 
\begin{equation}
\left|\frac{3\pi}{2}-\omega\right|  \leq \theta \leq  \pi \textrm{ with } \omega>\frac{\pi}{2} \label{intlimitspart}
\end{equation}
and to partial circles for
\begin{equation}
\left|\frac{\pi}{2}-\omega\right|  < \theta \leq  \left|\frac{\pi}{2}+\omega\right| \textrm{ with } \omega<\frac{\pi}{2} \textrm{ and }
\left|\omega-\frac{\pi}{2}\right|  < \theta \leq  \left|\frac{3\pi}{2}-\omega\right| \textrm{ with } \omega>\frac{\pi}{2} \textrm{ .} \label{intlimits}
\end{equation}
There are, however, always only full circles visible on either the day- or the nightside. The others are, together with the missing part of the partially visible circles, on the opposite half-sphere, turned away from the observer.

\subsection{Special case: Isotropic radiation}
If the observed spectra do not depend on the angle $\alpha$, we can perform the integration over the visible $\varphi$ and proceed analytically from eqn.~\ref{Asurface}. This is the case for isotropic emission from the surface and corresponds to a situation where the star shows neither limb-darkening nor limb-brightening. 
%Angles $\theta$ belong to fully visible circles on the dayside for $0\leq\theta\leq\left|\frac{\pi}{2}-\omega\right|$ with $\omega<\frac{\pi}{2}$, on the nightside for $\left|\frac{3\pi}{2}-\omega\right|\leq\theta\leq\pi$ with $\omega>\frac{\pi}{2}$ and to partially visible circles for $\left|\frac{\pi}{2}-\omega\right|<\theta\leq\left|\frac{\pi}{2}+\omega\right|$ with $\omega<\frac{\pi}{2}$ and $\left|\omega-\frac{\pi}{2}\right|<\theta\leq\left|\frac{3\pi}{2}-\omega\right|$ with $\omega>\frac{\pi}{2}$. There are, however, always only full circles on either the day- or the nightside visible. The others are, together with the missing part of the partial circles, on the opposite half-sphere.

For only partially visible circles the $\varphi$ integration goes up to $\alpha=\frac{\pi}{2}$, since there is always a 'grazing shot' when the circle moves out of sight, simplifying eqn.~\ref{eqnvarphi} to:

\begin{equation}
\varphi_1 = \arccos(-\cot{\theta}\cot{\omega})
\label{intlimitsphi}
\end{equation}
%For $\theta$ the limit of integration is $(\omega-\frac{\pi}{2})\leq\theta\leq(\omega+\frac{\pi}{2})$ (since the observer angle can always be turned in the $\varphi = 0$ plane), while for $\varphi$ the borders are given by Eqn. (\ref{intlimitsphi})

We decompose eqn.~\ref{Atotal} in separate integrals for full (day- or night-side) $A_{\mathrm{f}}$ and partial circles $A_{\mathrm{p}}$. Depending on the values of $\theta_{1}$ and $\theta_{2}$ one of them or a combination of both is used according to the limits on $\theta$ given in eqns.~\ref{intlimitsfull}-\ref{intlimits}.

\begin{eqnarray}
A_{\mathrm{f}} &= & 2\int^{\theta_{2}}_{\theta_{1}} \int^{\pi}_{0} R^{2} \sin{\theta} (\sin{\theta}\cos{\varphi}\sin{\omega} + \cos{\theta}\cos{\omega}) \mathrm{d}\varphi \; \mathrm{d}\theta =  \left. R^{2} 2\pi \cos{\omega} \; \frac{1}{2} \sin^{2}{\theta} \; \right|^{\theta_{2}}_{\theta_{1}}\\
A_{\mathrm{p}} &= &2\int^{\theta_{2}}_{\theta_{1}} \int^{\varphi_1}_{0} R^{2} \sin{\theta} (\sin{\theta}\cos{\varphi}\sin{\omega} + \cos{\theta}\cos{\omega}) \mathrm{d}\varphi \; \mathrm{d}\theta \nonumber \\
& = & \int^{\theta_{2}}_{\theta_{1}} R^{2} \sin^{2}{\theta} \sin{\omega} 2 \sqrt{1 - \cot^{2}{\theta} \cot^{2}{\omega}} + R^{2} \sin{\theta} \cos{\theta} \cos{\omega} 2 [\pi - \arccos (\cot{\theta}\cot{\omega})] \; \mathrm{d}\theta \label{I2expression}
\end{eqnarray}
with the upper limit on $\varphi$ from eqn.~\ref{intlimitsphi}. The $\theta$ integration in Eqn. (\ref{I2expression}) still has to be solved numerically.

\subsection{Coverage}
The problem is more complicated in the case of eclipsing binaries. In this case the eqns.~\ref{intlimitsfull}-\ref{intlimits} need to be extended, so that areas, which are covered by the primary are excluded from the integration.
As one can see in Fig. \ref{fig:geom} (right panel) the first latitudes to be affected are full circles seen close to grazing-incidence.

To describe this we need a $\varphi(\theta)$ that uses radii, distance and angle in the plane of motion to adjust the integration limits accordingly.

There are two special cases, one for a primary larger than the secondary, where a full eclipse might occur during transit and one for a primary smaller than secondary, where the the primary only covers the center of the secondary and hence four integration limits are necessary to describe both visible sides during transit.

In short, we parametrise the edge of the primary as projected on the secondary in the plane of the sky with $x', y', z'$ for a circle:
\begin{equation}
y'=\sqrt{r_p^2-(x'+R\sin{\omega})^2} \label{eqnydash}
\end{equation}
with $r_p$ radius of the primary and $R$ separation between both measured from centre to centre.

Then we transform this system with a rotation around the $y$-axis to our old set of coordinates $x, y, z$ in eqn.~\ref{r}:
\begin{equation}
\left(\begin{array}{c}
x' \\
y' \\
z'
\end{array}\right)
=
\left(\begin{array}{c}
\sin{\theta} \cos{\varphi} \cos{\omega} + \cos{\theta} \sin{\omega} \\
\sin{\theta} \sin{\varphi} \\
\cos{\theta} \cos{\omega} - \sin{\theta} \cos{\varphi} \sin{\omega}\\
\end{array}\right) \label{eqndashed}
\end{equation}

Substituting all dashed coordinates in eqn.~\ref{eqnydash} with eqn.~\ref{eqndashed} we get the following expression:
\begin{equation}
\sin{\theta}\sin{\varphi}=\sqrt{r_p^2-([\cos{\theta}+R]\sin{\omega}+\sin{\theta}\cos{\varphi}\cos{\omega})^2}
\end{equation}
%Since in this case $\omega < 0$ and $\sin(-\omega) = -\sin(\omega)$ we get a change in sign for the second term under the square root.

Depending whether the $\theta=\textrm{const}$ ring is still a full circle, a partial circle or  the primary covers the substellar point we get new integration limits, still taking into account eqns.~\ref{intlimitsfull}-\ref{intlimits}.

For the outer  marked ring shown in the example of Fig.~\ref{fig:geom} (right panel) the new limits are $\varphi_{cover, lower}=\frac{2}{9}\pi$ and $\varphi_{cover, higher}=\pi$ respectively.

\subsection{Phase and inclination}

We still need to calculate $\omega$ from the phase $p$ and the inclination
$i$. This can be done with the scalar product between the line-of-sight and the $z$-axis:
\begin{equation}
\left(\begin{array}{c}
\cos{p}\cos{i} \\
\sin{p} \\
\cos{p}\sin{i}
\end{array}\right)
\left(\begin{array}{c}
0 \\
0 \\
1
\end{array}\right)
= \cos{p} \sin{i} = \cos{\omega} 
\quad \Longrightarrow \quad \omega = \arccos(\cos{p} \sin{i})
\end{equation}

\subsection{IDL code}
We provide an IDL code, that implements the formulae derived in this section in electronic form at the CDS via anonymous ftp to cdsarc.u-strasbg.fr (130.79.128.5) or at \texttt{http://www.hs.uni-hamburg.de/DE/Ins/Per/Guenther/codes/geom.pro}~.

\section{A star in 1.5 D}
\label{star15d}
Our geometric derivation is independent of the mechanism, which causes the different zones on the star. In this section we provide some examples choosing a specific spectral synthesis code to illustrate the use case of the formulae above.

\subsection{Simulation setup}
In the following we show simulated spectra of irradiated stars. The spectra are calculated using a modified version of the PHOENIX~15.04 stellar atmosphere code \citep{1999JCoAM.109...41H}. It can consistently produce the incident flux spectrum from the hot sdO \citep{2001PASP..113..119A} and model the cooler MS companion \citep{2001ApJ...556..357A}. It also includes an irradiation mode \citep{2004ApJ...614..338B,2005ApJ...632.1132B}, which treats the radiation from a primary star in the outer boundary condition of the radiation transport equation. H and He are treated in full NLTE.
%The most important elements for our analysis (H, He, C, N and O) are treated in full NLTE.

We combine 1D models calculated for different inclination angles and compare the resulting spectra. We present two examples at very different temperatures. 

\subsection{Limitations of the model}
PHOENIX solves the hydrodynamic equations in 1D, thus, for each zone on the secondary, where we calculate a spectrum, the hydrodynamical structure adjusts according to the energy input from the primary star. Therefore, the temperatures, pressures and densities in each layer differ between the separate zones we model. This should lead to horizontal energy transport between the zones in the form of heat conduction or winds, but our model cannot account for that. For small gradients and large physical dimensions this is an acceptable approximation. It could be improved by matching the entropy of the different rings in the convection zone to account for horizontal mixing in this layer \citep{1993MNRAS.264..641B,2001MNRAS.327..989C}. \citet{2004ApJ...614..338B} used this method for pre-CVs. However, we decided to keep the energy input at the lower boundary fixed according to the effective temperature of the non-irradiated secondary. Thus, the irradiated zones do not reach the same adiabat at depth. Entropy-matching requires a grid of models with different intrinsic effective temperatures for each irradiation angle, which is beyond the scope of the simple example we want to provide here. Because we enforce radiative equilibrium throughout the model, the total albedo is equal to 1.0 \citep{1926MNRAS..86..320E}.

\subsection{Irradiation by an exceptionally hot primary}
First, we choose a situation which closely resembles a pre-CV composed of a hot primary sdO and a secondary MS star. Reviews about detached binaries, physical processes in close binary systems, and general three-dimensional fluid dynamics in binary systems are given by \citet{2001LNP...563....1C}, \citet{2002MNRAS.335..358B}, and \citet{2000NewAR..44..119M}.
The setup is similar to the \object{UU Sge} system \citep{1966ApJ...144..259A,1978ApJ...223..252B,1993MNRAS.262..377P} as modelled in \citet{2009A&A...505..227W}, but here we show results for an extreme situation with a much hotter primary with $T_p=125\;000$~K, which irradiates a main-sequence secondary. The centres of mass of both components are separated by 2.5~R$_{\sun}$, the radius of the primary is 0.34~R$_{\sun}$, and  the radius of the irradiated secondary is 0.27~R$_{\sun}$.

Due to the extrem irradiation conditions the models are numerically fragile. In the simulations we present the temperature correction between the iterations is at most a few~\%, indicating that the models are close to the physical equilibrium state. %However, for a larger number of iterations, they do not converge further but the numerical instabilities drive the models to unphysical values.

\begin{figure}
\centering
\resizebox{0.48\hsize}{!}{\includegraphics{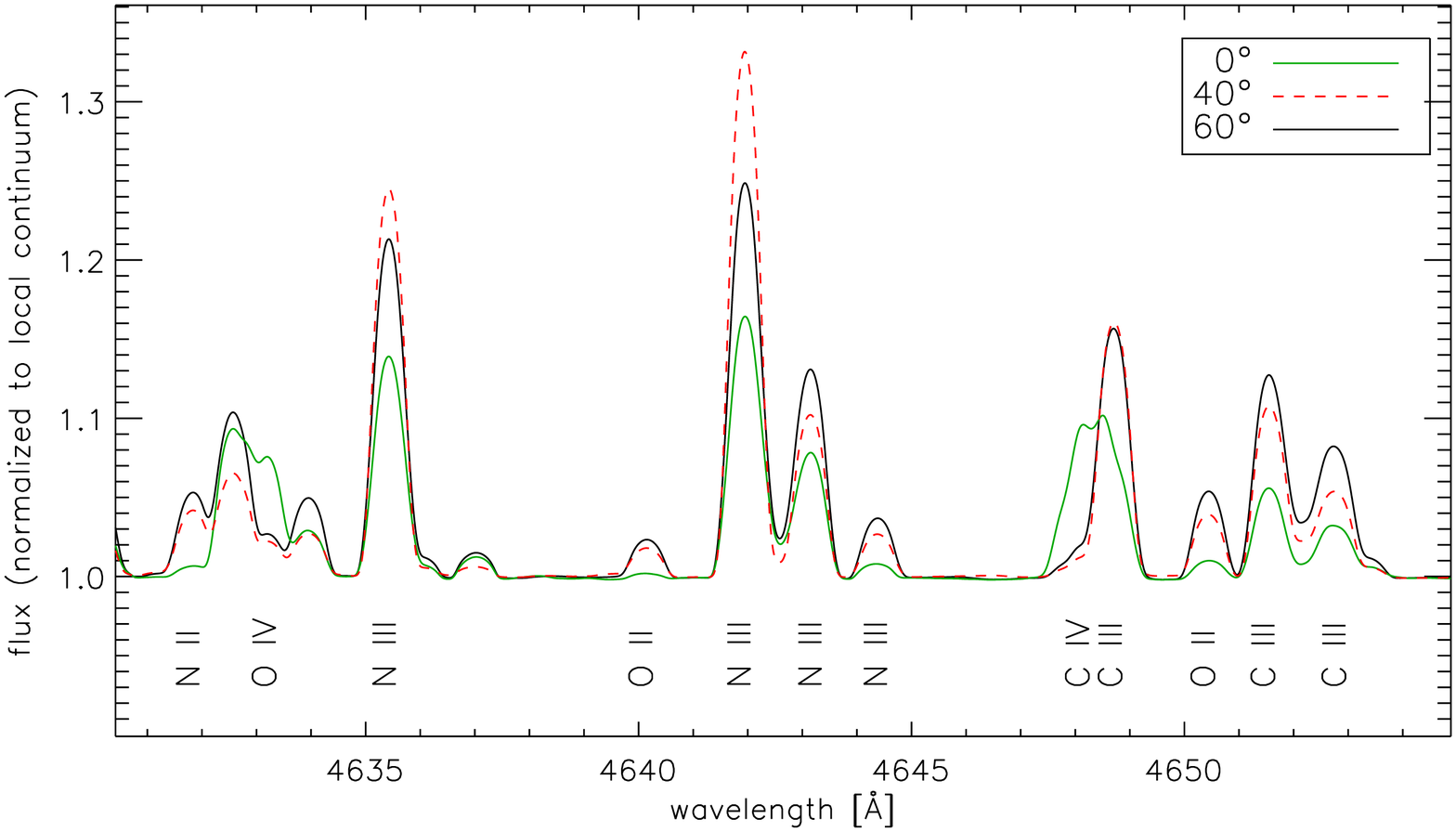}}
\resizebox{0.48\hsize}{!}{\includegraphics{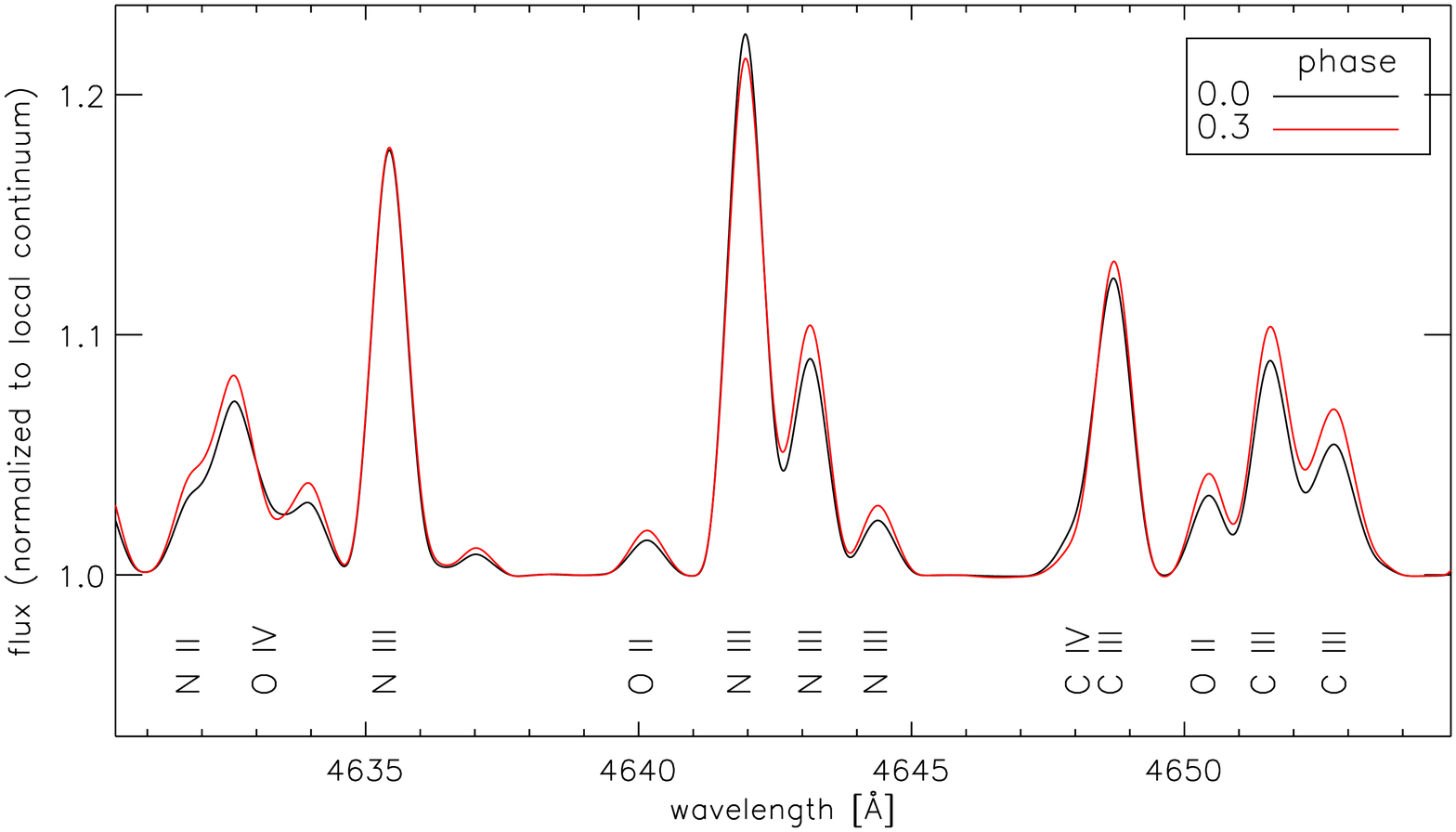}}
\caption{\emph{left:} Spectra for three different irradiation angles with irradiation by a close hot primary. The lines of C, N and O are labelled. (Colour in electronic version only) \emph{right:} Combined spectrum for different phases and inclination $i=85\degr$ with irradiation by a close hot primary. (Colour in electronic version only) \label{fig:irrad_spec} }
\end{figure}

In Fig.~\ref{fig:irrad_spec} (left panel) we show spectra for three different irradiation angles (measured towards the normal of the stellar surface). They are all normalised to their respective local continuum. Several species can be seen in emission, because the atmosphere is heated from the outside. The emission lines from different ionisation stages of C, N and O are labelled. 
Conventionally, a single temperature is used as effective temperature of the heated hemisphere, but the figure shows that no single temperature can exist to fit a combination of these three spectra. The \ion{N}{ii} line at 4632~\AA{} is strongest for large irradiation angles, that is atmospheres with a comparatively low temperature, but the strong \ion{N}{iii} lines at 4635~\AA{} and 4642~\AA{} are most pronounced for medium irradiation angles. For oxygen the region we show includes higher states of ionisation. As in the case of nitrogen, the \ion{O}{ii} lines are strongest in the model with an irradiation angle of 60$^{\circ}$, whereas \ion{O}{iv} at 4633~\AA{} is strongest for the model calculated at the substellar point, where the irradiation is strongest. The same can be observed for \ion{C}{iii}. The \ion{C}{iii} line at 4649~\AA{} shows the same flux for the models with irradiation angle 40$^{\circ}$ and 60$^{\circ}$, in contrast to the lines at 4652~\AA{} and 4653~\AA{}.
% \begin{figure}
% \centering
% \resizebox{\hsize}{!}{\includegraphics{total_spec}}
% \caption{Combined spectrum for different phases and inclination $i=5\degr$ with irradiation by a close hot primary. (Colour in electronic version only) \label{fig:spectrum_comb} }
% \end{figure}
The three spectra from Fig.~\ref{fig:irrad_spec} (left panel) are then combined to form a stellar spectrum integrated about the stellar disk according to eqns.~\ref{Atotal}-\ref{I2expression}. The zones we use cover the area $\theta=0\degr-20\degr$, $20\degr-50\degr$ and $50\degr-90\degr$ (green, red and black spectrum in Fig.~\ref{fig:irrad_spec}, left panel). Their effective temperatures are 41\,000~K, 35\,000~K and 32\,000~K. As an example we combine those spectra to calculate the emission for an inclination $i=85\degr$ and phases 0 and 0.3. The relative area of these three zones is 0.15, 0.46 and 0.39 for phase 0  according to the equations in Sect.~\ref{geometry}. With increasing phase the sub-stellar point rotates out of view and at phase 0.3 only the outer regions play a role with a relative area of 0.15 ($20\degr-50\degr$) and 0.85 ($50\degr-90\degr$).  The contribution to the emission lines from the night-side is negligible because its effective temperature is only $3\;400$~K. Due to the irradiation the substellar point is the brightest part of the photosphere and when it rotates out of view, the observed luminosity of the secondary decreases. Figure~\ref{fig:irrad_spec} (right panel) shows the predicted lines normalised to the local continuum. For phase 0.3 those lines, which are most prominent for large $\theta$ values, e.g. the \ion{O}{ii} and \ion{C}{iii} line between 4650~\AA{} and 4654~\AA{}, are stronger than for phase 0 relative to the continuum. The spectrum contains lines of all ionisation stages discussed above from \ion{N}{ii} to \ion{C}{iv}. Thus, only a model which distinguishes different zones on the day-side can predict the changes of the emission lines with the phase of the binary.
\subsection{Irradiation by a cooler white dwarf}
\begin{figure}
\centering
\resizebox{0.48\hsize}{!}{\includegraphics{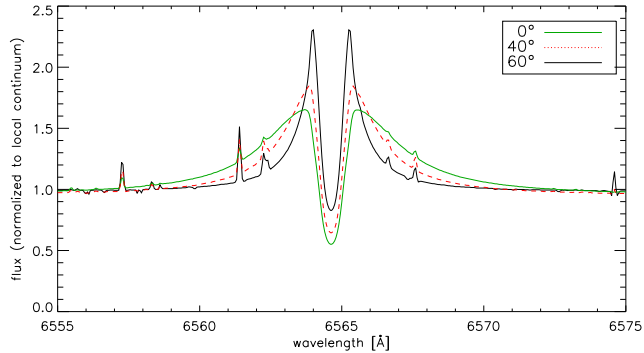}}
\caption{H$\alpha$ line profiles for three different irradiation angles from a close sdO primary. (Colour in electronic version only) \label{fig:ha_spec} }
\end{figure}

In this second example, we irradiate the secondary with the spectrum of a primary with effective temperature of only $20\;000$~K. The intrinsic temperature of the secondary is $4\;100$~K. All other parameters are similar to the first example:  The centres of mass of both components are separated by 2.5~R$_{\sun}$, the radius of the primary is 0.34~R$_{\sun}$, and 0.53~R$_{\sun}$ for the irradiated secondary ($T_{\mathrm eff}=4100$~K on the night-side). We use the same $\theta$ boundaries as above. Their effective temperatures are 6500~K, 6000~K and 5400~K. The lower irradiation intensity leads to better numerical stability in the simulations. The atmosphere of the secondary is relativley cool and does not form highly ionised ions. The most notable difference between the rings with different irradiation angles can be seen in the H$\alpha$ line profile (Fig.~\ref{fig:ha_spec}), which is broad for the sub-stellar point, where the irradiation is more intense.
Again, changes in the line profile occur when the system rotates with respect to the observer.

\begin{figure}
\centering
\resizebox{0.48\hsize}{!}{\includegraphics{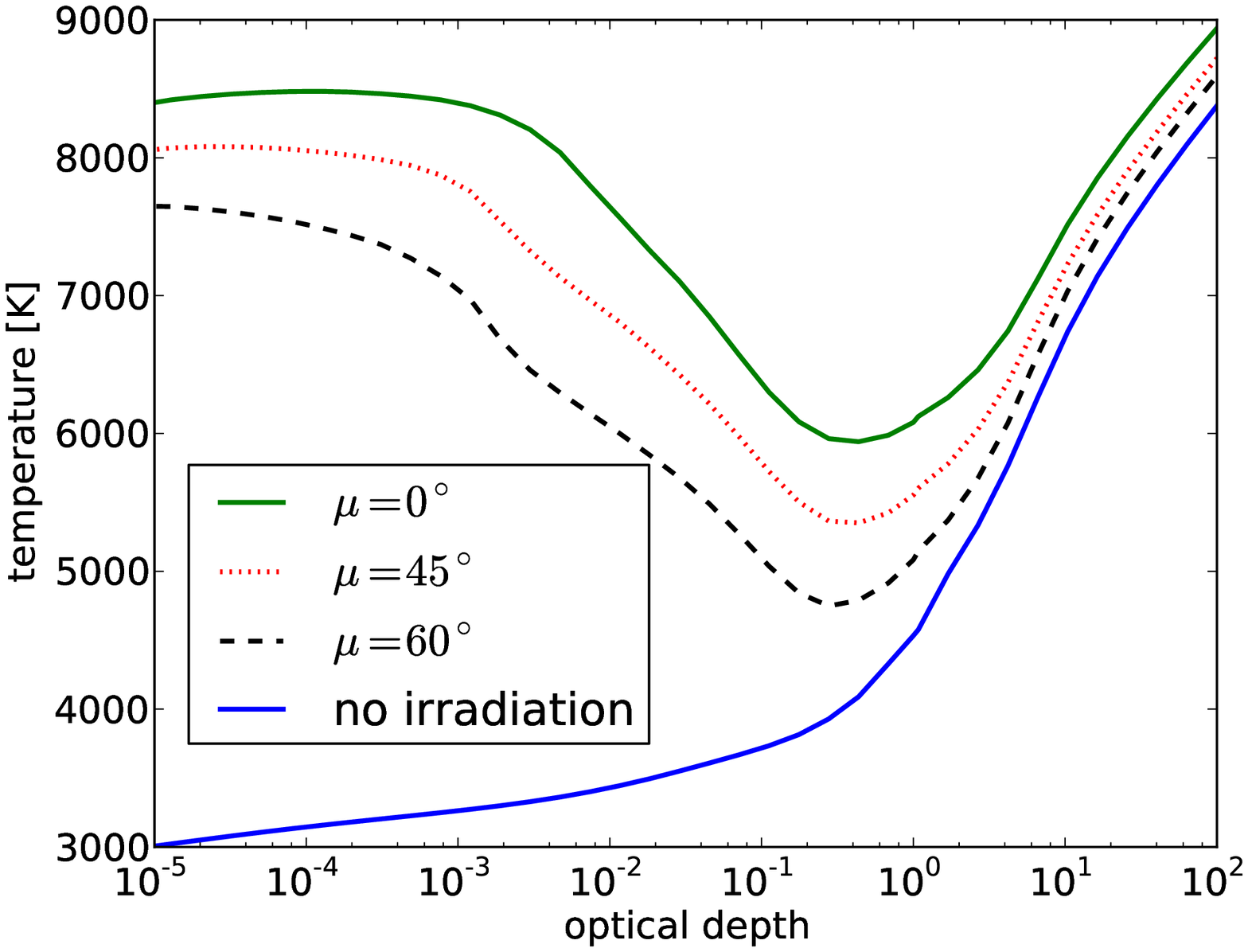}}
\resizebox{0.48\hsize}{!}{\includegraphics{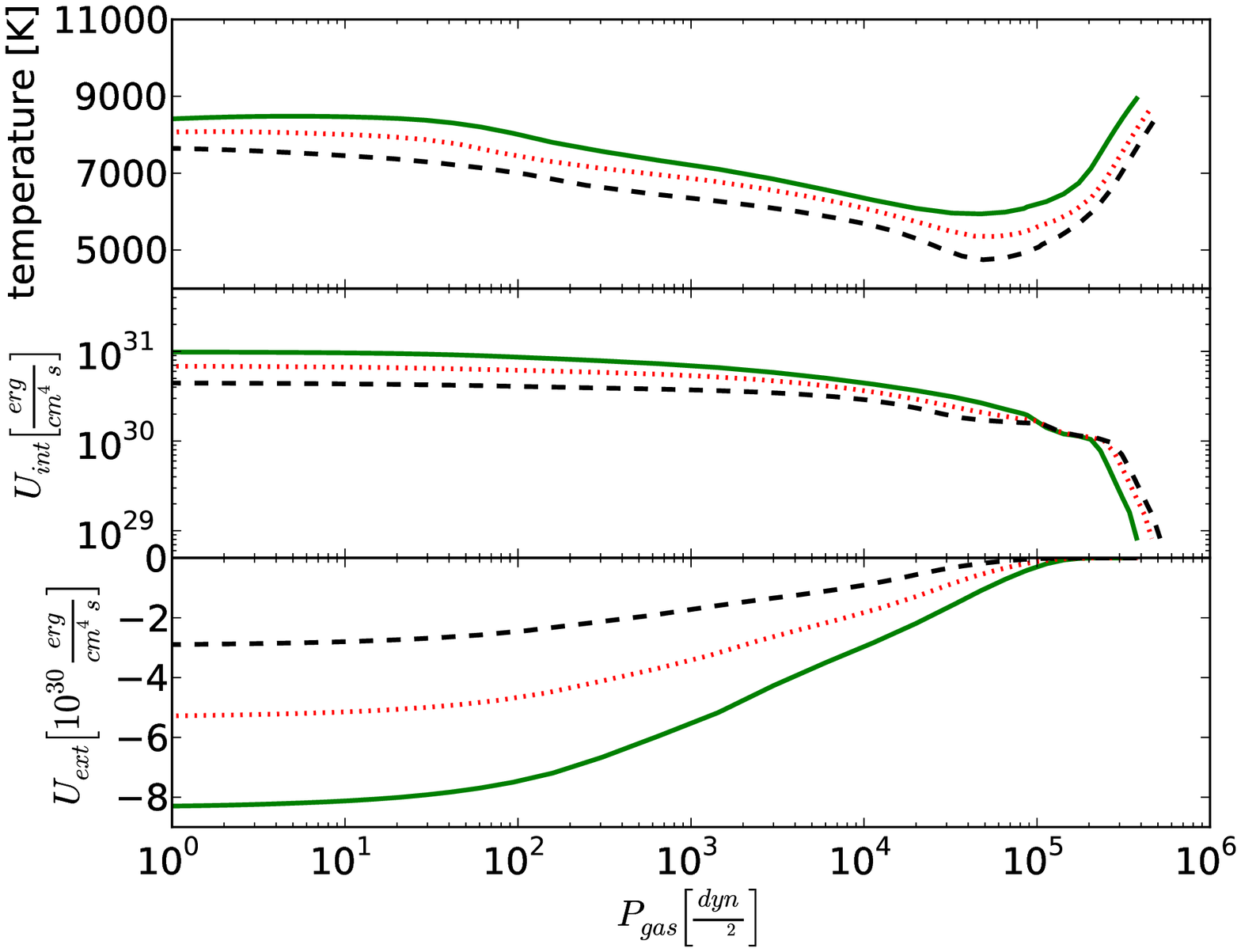}}
\caption{\emph{left:} Temperature of optical depth for the different zones of a secondary irradiated by a cool white dwarf (see text for details of the model). The lines show the following situations (top to bottom): Irradiation at the substellar point (green), at $\mu = 45\degr$ (dotted, red), $\mu = 60 \degr$ (dashed, black) and the night-side, i.e. no irradiation (blue). (Colour in electronic version only) \emph{right:} The structure of the temperature and radiation field in the irradiated atmospheres. The upper panel shows electron temperature $T_{{\rm elec}}$, the middle panel the internal radiation flux $u_{{\rm H,~int}}$, and the lower panel the external radiation flux $u_{{\rm H,~ext}}$. See text for details of the definition of the fluxes. The different zone are marked as in the left panel. (Colour in electronic version only) \label{fig:temptau} }
\end{figure}

For this example we show the effect of the irradiation on the secondary's atmosphere in more detail in Fig.~\ref{fig:temptau}. The left panel shows the evolution of the temperature with the optical depth $\tau$, which is evaluated as the continuum opacity at $1.2 \mu $m. The atmosphere in the zone, which represents the sub-stellar point, is the hottest for each optical depth. The temperatures decrease with decreasing incident flux. All three irradiated models show a temperature-inversion with high temperatures on the outside and a temperature minimum close to $\tau = 0.5$. In deeper layers the temperature rises again and then follows the unirradiated model for the same stellar parameters. In the layers before the temperature minimum ($\tau = 10^{-3}-0.5$) the optical depth of the models increases with increasing irradiation for a given column density of the atmosphere, thus the irradiation increases the opacity of the gas. This effect has been observed first by \citet{1993MNRAS.264..641B}. 
The right panel of Fig.~\ref{fig:temptau} describes the structure on the atmosphere in relation to the gas density. For comparison purposes the upper panel again shows the evolution of the temperature. For all irradiated models the temperature inversion occurs around $P_{gas} = 10^5$~dyn~cm$^{-2}$. Our models are calculated to gas temperatures about one order of magnitude larger than that. 

The remaining two panels of Fig.~\ref{fig:temptau} characterise the radiation field: The middle panel shows the internal radiation flux $u_{{\rm H,~int}}$, defined as the radiation that an atmosphere with the temperature structure from the upper panel emits in the absence of external irradiation. The inner boundary condition matches $u_{{\rm H,~int}}$ to the blackbody radiation for the intrinsic, i.e., undisturbed, temperature of the secondary.

$u_{{\rm H,~int}}$ is a net flux, thus all inward radiation, which is reflected at the inner boundary, cancels out. In contrast, the radiation going outwards is summed up and therefore $u_{{\rm H,~int}}$ increases monotonically outwards by about two orders of magnitude. The emitted energy of all layers exceeds the energy delivered from the inner boundary, so the atmosphere would cool down in the absence of an additional energy source until $u_{{\rm H,~int}}$ matches the intrinsic flux at the inner boundary for all layers. At low pressures $u_{{\rm H,~int}}$ is larger for irradiation at smaller $\mu$ because these zones are hotter in general. The differences at the high density are due to the setup of the model. All models are calculated to a constant value of $\tau$, which corresponds to different gas pressures for the different models. Thus, the inner boundary, where $u_{{\rm H,~int}}$ is held fixed, is reached at different gas pressures.

The lower panel shows the external radiation flux $u_{{\rm H,~ext}}$. Here, a full model including the internal flux and the irradiation is calculated. The difference between the radiation field obtained in this case and the internal radiation flux ($u_{{\rm H,~int}}$) from the previous panel is $u_{{\rm H,~ext}}$. It is negative, because the flux is directed inwards.

The plot does not contain reflected external irradiation, since this cancels out in the net flux $u_{{\rm H,~ext}}$. The irradiation per unit surface area is larger for points with small $\mu$ because they are closer to the surface of the primary and the irradiation angle is steeper. In the outer layers of the atmosphere the optical depth is low, so only little flux is absorbed. Most of the heating occurs for gas pressures of $10^2-10^5$~dyn cm$^{-2}$. At the bottom of this zone the external radiation field is completely absorbed and consequently the temperature inversion occurs. For small values of $\mu$ the radiation penetrates slightly deeper into the atmosphere, because the path length to reach the same depth is shorter than for shallower incident angles.

\section{Summary and Conclusion}
\label{conclusion}
In a binary system both components may influence each other. We set up a coordinate system, where the z-axis connects the centres of mass of both components and concentrate on the secondary. Assuming that the influence of the primary can be parameterised with the angle $\theta$ between surface normal and z-axis we calculate the area of a ring between two $\theta$ values projected on the plane of the sky of an observer \emph{at arbitrary viewing position}, this is meant to be used as weighting factor when synthesizing spectra. The formalism includes occultation of the secondary in an eclipse if appropriate. The pre-requisites are purely geometrical and several effects, which change the  spectrum of the secondary, can in principle be treated according to our formalism: Irradiation, gravitation (as long as the secondary stays spherical), abundance anomalies, wind impact.

We calculate synthetic spectra with the PHOENIX stellar atmosphere code for two different irradiating primaries, assuming negligible horizontal heat transport, so that rings of constant temperature develop around the substellar point. The spectra change with the irradiation angle, thus a simulation combining multiple zones is more accurate than representing the entire day-side with a single temperature component. We showed that simulations with multiple zones can be combined from established present-day stellar atmosphere models. These calculations provide a benchmark for future full 3D models, which should reproduce our results in the limit of negligible horizontal heat flux.

\begin{acknowledgements}
The authors thank T.~Barman for his support in setting up calculations of irradiated atmospheres with PHOENIX. H.M.G. acknowledges support from DLR for project number 50OR0105. ACW was supported by the DFG (Deutsche Forschungsgemeinschaft), project number HA 3457/7-1. Calculations were performed at the Hamburger Sternwarte Delta Opteron Cluster ('Nathan') financially supported by the DFG and the State of Hamburg.
\end{acknowledgements}

\bibliographystyle{../../aa-package/bibtex/aa} % style aa.bst
\bibliography{../articles}
\end{document}